\documentclass[12pt,fleqn,a4paper]{article}
\usepackage{amssymb}
\usepackage{amsmath}
\usepackage[dvips]{graphicx,psfrag}
\begin{document}
\begin{flushright}
TU-650 \\
{\tt hep-ph/0204164 }\\
\end{flushright}
\vspace*{1.5cm}
\begin{center}
    {\baselineskip 25pt
    \Large{\bf 
    
    Small Dirac Neutrino Masses \\
    in Supersymmetric Grand Unified Theories
    
    }
    }

    \vspace{1.2cm}
    \def\thefootnote{\fnsymbol{footnote}}
    {\large Ryuichiro Kitano}\footnote
    {email: {\tt kitano@tuhep.phys.tohoku.ac.jp}}

    \vspace{.5cm}
    
    {\small {\it 
    Department of Physics, Tohoku University, Sendai 980-8578, Japan
    }}
    
    \vspace{.5cm}
    \today
    
    \vspace{1.5cm}
    {\bf Abstract}

\end{center}

\bigskip

A simple mechanism to generate Dirac masses
for the neutrinos in SU(5) supersymmetric grand unified
theory is proposed.
The tiny Dirac masses are induced by
the small mixing between the Higgs fields
and another superheavy fields.
The mixing terms are obtained by the 
same mechanism as the $\mu$-term generation
of the order of the supersymmetry breaking scale,
so that the mixing of order TeV$/M_{\rm GUT} \sim 10^{-13}$
is realized.
We consider the lepton flavor violating processes
in this model.
The branching ratios are directly related to
the neutrino oscillation parameters
and we can predict 
the $B(\tau \to \mu \gamma) / B(\mu \to e \gamma) $ ratio
once the neutrino oscillation parameters are determined.


\newpage
\def\thefootnote{\arabic{footnote}}
\setcounter{footnote}{0}
\baselineskip 20pt


The data from SuperKamiokande suggest 
the presence of tiny neutrino masses which
clearly indicates a necessity of an
extension of the lepton sector 
in the minimal Standard Model \cite{Fukuda:1998mi}.
The easiest extension
is to introduce the right-handed neutrinos
and the Yukawa interaction terms
such as $f_\nu^{ij} (\bar{l}_i h) \nu_{Rj}$,
where $l$, $h$, and $\nu_R$ are the left-handed lepton doublets,
the Higgs field and the right-handed neutrinos, respectively.
These terms induce the Dirac masses for the neutrinos
through the SU(2)$_L$ $\times$ U(1)$_Y$ breaking effect.
In that case,
the coupling constants $f_\nu$ must be very small of order $10^{-13}$
in order to reproduce the tiny neutrino masses.
Also, the neutrinos can acquire Majorana masses
by adding terms as $\lambda_{ij} (\bar{l}_i h) (l_j h^\dagger)$.
The parameters $\lambda_{ij}$ have mass dimension of $-1$ and
should be of the order of $10^{-14}$ GeV$^{-1}$.
In the framework of the seesaw mechanism,
the smallness of $\lambda_{ij}$ is explained by
large Majorana masses of the right-handed neutrinos 
$M_R \sim 10^{14}$ GeV \cite{seesaw}.

In any case,
we need somewhat unnatural parameters $f_\nu \sim 10^{-13}$
or $M_R \sim 10^{14}$ GeV.
In Supersymmetric (SUSY) Grand Unified Theories (GUTs),
there is a hint toward this problem.
The SUSY GUT provides
scales of $M_{\rm SUSY} \sim$ TeV 
and $M_{\rm GUT} \sim 10^{16}$ GeV
to the theory,
while
the ratio $M_{\rm SUSY} / M_{\rm GUT}$
is adequate for the Dirac Yukawa couplings $f_\nu \sim 10^{-13}$.
Several attempts to utilize the SUSY breaking scale 
have been made \cite{Arkani-Hamed:2000bq,Borzumati:2000fe}.
Especially, in ref.\cite{Borzumati:2000fe},
Borzumati {\it et al.} considered the possibility 
of explaining the LSND data \cite{Athanassopoulos:1996wc}
by introducing a sterile neutrino field ($S$)
whose mass is given by the SUSY breaking effect
at the suitable order ($M_{\rm SUSY}^2 / M_R \sim 1$ eV).
The Dirac mass terms between $S$ and the right-handed neutrinos $N$
are forbidden by $R$-symmetry at first,
and given through a non-renormalizable
interactions of $ \langle W \rangle S N / M_{\rm Pl}^2 \sim 
m_{3/2} S N$ where $W$ and $m_{3/2}$ are the superpotential
and the gravitino mass, respectively.
In supergravity scenario \cite{Hall:1983iz},
the vacuum expectation value (VEV) of $W$ is necessary 
to cancel the cosmological constant caused 
by the SUSY breaking sector.
They also pointed out 
the existence of the Dirac mass terms 
$ \langle W \rangle L H S / M_{\rm Pl}^3 $.
Although the Dirac Yukawa coupling constants 
from those terms are 
too small of order $m_{3/2} / M_{\rm Pl} \sim 10^{-15}$,
it is interesting that the replacement of $M_{\rm Pl}$ to
$M_{\rm GUT}$ gives suitable magnitude for the neutrino oscillation.

Another type of the Dirac neutrino scenario have been proposed 
by Mohapatra and Valle in superstring models \cite{Mohapatra:bd}.
The small Dirac neutrino masses are obtained
by hierarchy between VEVs of 
two distinct standard model singlet components in
${\bf 27}$ and ${\bf \bar{27}}$ representations
of the ${\rm E_6}$ group.

In this paper,
we propose a mechanism to generate tiny Dirac Yukawa couplings
$f_\nu$ in the context of SU(5) SUSY GUT.
The mechanism
is similar to the usual seesaw scenario
or the Froggatt-Nielsen mechanism \cite{Froggatt:1978nt}.
In the Froggatt-Nielsen mechanism,
the small mass parameters are explained 
by imposing U(1) symmetry
and introducing a small VEV
of a U(1) breaking field.
In our case, as in ref.\cite{Borzumati:2000fe},
we use $R$-symmetry as such U(1) symmetry
and the small $R$-symmetry breaking terms
are automatically supplied by the SUSY breaking effect
as in the $\mu$-term generation mechanism
\cite{Hall:1983iz,Giudice:1988yz,Casas:1993mk}.
By using the above mechanism,
the tiny coupling of order $10^{-13}$ is naturally obtained 
as the ratio $M_{\rm SUSY} / M_{\rm GUT}$
through a mixing between the usual Higgs doublet and 
another superheavy Higgs doublet.
We also consider the Lepton Flavor Violating (LFV) processes
as a low energy prediction of our model.

We construct an SU(5) GUT model.
We assign $R$-charges for the matter fields as follows:
\begin{eqnarray}
 \bar{F}: ({\bf \bar{5}}, 1)\ ,\ \ \ 
 T: ({\bf 10}, 1)\ ,\ \ \ 
 N: ({\bf 1}, -1)\ ,
\end{eqnarray}
where the former and latter numbers are the dimension 
of the representation of the SU(5) group and 
the $R$-charge, respectively.
The superfields $\bar{F}$ and $T$ represent the usual matter
and $N$ is the right-handed neutrino superfield.
In the Higgs sector, in addition to the usual
Higgs fields, we introduce a pair of 
${\bf 5}$ and ${\bf \bar{5}}$ representation fields $H^\prime$
and $\bar{H}^\prime$ as follows:
\begin{eqnarray}
 H: ({\bf 5}, 0)\ ,\ \ \ 
 \bar{H}: ({\bf \bar{5}}, 0)\ ,\ \ \ 
 H^\prime: ({\bf 5}, 2)\ ,\ \ \ 
 \bar{H}^\prime: ({\bf \bar{5}}, 0)\ .
\end{eqnarray}
The superpotential relevant to the mechanism is written as follows:
\begin{eqnarray}
 W = \tilde{f}_\nu^{ij} \bar{F}_i H^\prime N_j
   + M_{H^\prime} H^\prime \bar{H}^\prime \ .
\label{3}
\end{eqnarray}
The $H^\prime \bar{H}$ term which is allowed by
$R$-symmetry can be eliminated by the field redefinition of
$\bar{H}$ and $\bar{H}^\prime$.
The Yukawa coupling constants $\tilde{f}_\nu$ and 
the mass parameter $M_{H^\prime}$ are naturally taken to be 
of order unity and the GUT scale of $10^{16}$ GeV, respectively.

The essential point is that the SUSY breaking effect 
induces $R$-symmetry breaking (but $R$-parity conserving) terms
for the combination of vanishing $R$-charge
such as 
\begin{eqnarray}
 W_{R{\rm -breaking}} = \mu H \bar{H} + \mu^\prime H \bar{H}^\prime\ ,
\end{eqnarray}
with the mass parameter $\mu$ and $\mu^\prime$ of the order of
SUSY breaking scale such as TeV.
These terms are naturally induced  
by the Giudice-Masiero mechanism \cite{Giudice:1988yz}
in the supergravity scenario,
and other mechanisms have been considered
in the literature \cite{Hall:1983iz,Casas:1993mk}.
The tiny Dirac neutrino masses can be obtained
with such terms.
The existence of $\mu^\prime$ term induce the 
tiny mixing $\delta_{H,H^\prime}$ 
between $H$ and $H^\prime$ fields
as $\delta_{H,H^\prime} \sim \mu^\prime / M_{H^\prime} $.
It follows that 
the low energy effective superpotential
are described in terms of the light Higgs multiplets 
$H_{l}$ and $\bar{H}_l$,
which are almost $H$ and $\bar{H}$,
as follows:
\begin{eqnarray}
 W_{\rm eff} = \delta_{H,H^\prime} \tilde{f}_{\nu}^{ij}
 \bar{F}_i H_{l} N_j + \mu H_l \bar{H}_l \ .
\end{eqnarray}
The first term is the usual Dirac Yukawa interaction
with coupling constants 
$f_\nu = \delta_{H,H^\prime} \tilde{f}_\nu \sim 10^{-13}$
which is suitable for the neutrino oscillation data.

In general,
if we assume the supergravity scenario,
the small Majorana mass terms for $N$ may be induced by
$\langle W \rangle^2 N^{2} / M_{\rm Pl}^5 $ term
of the order of $m_{3/2}^2 / M_{\rm Pl} \sim 10^{-3}$ eV
\cite{Borzumati:2000fe}.
If we include this contribution,
we have a possibility to realize the pseudo-Dirac scenario
\cite{Wolfenstein:1981kw},
where the large mixing of the neutrinos and 
the data from the LSND experiment \cite{Athanassopoulos:1996wc}
can be naturally explained.
However, 
the data from SNO experiments 
support the solar neutrino oscillation of $\nu_e$ to
an active neutrino \cite{McGregor:2002sc},
so that the Majorana masses for $N$ are strongly constrained.
In order to escape this constraint,
we need to assume that the neutrinos do not directly couple to
the cosmological constant tuning sector.
Another possible way to avoid small Majorana mass terms for $N$ is
imposing global $B-L$ like symmetry
with the charge of $\bar{F}: -3$, $T: 1$, $N: 5$,
$H: -2$, $\bar{H}: 2$, $H^\prime : -2$, and $\bar{H}^\prime: 2$.
Also, in the gauge mediated SUSY breaking scenario
\cite{Dine:1993yw},
the contribution is negligible.
Similarly, as mentioned before,
the Dirac mass terms
$\langle W \rangle \bar{F} H N / M_{\rm Pl}^3
\sim 10^{-15} \bar{F} H N$
may arise \cite{Borzumati:2000fe}
in the supergravity scenario.
These contributions to the Dirac masses are negligible compared 
to those from the above Higgs mixing effects.

\begin{figure}[t]
\hspace*{2.5cm}
\includegraphics[width=12cm]{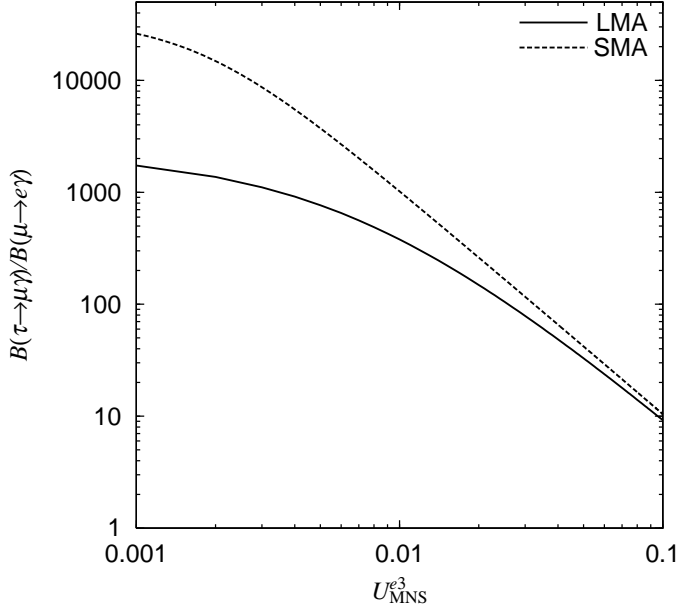}  
\caption{
The ratio of $B(\tau \to \mu \gamma) / B(\mu \to e \gamma)$
is plotted.
The horizontal axis is $U^{e3}_{\rm MNS}$.
The solid and dashed lines represent
the large and small angle MSW solution to the solar neutrino
problem, respectively.
\label{fig1}
}
\end{figure}
\begin{figure}[th]
\begin{center}
\psfrag{a}{\small $10^{-11}$}
\psfrag{b}{\small $10^{-12}$}
\psfrag{c}{\small $10^{-13}$}
\psfrag{d}{\small $10^{-14}$}
\psfrag{e}{$m_0$ (GeV)}
\psfrag{f}{$M_{1/2}$ (GeV)}
\includegraphics[width=10cm]{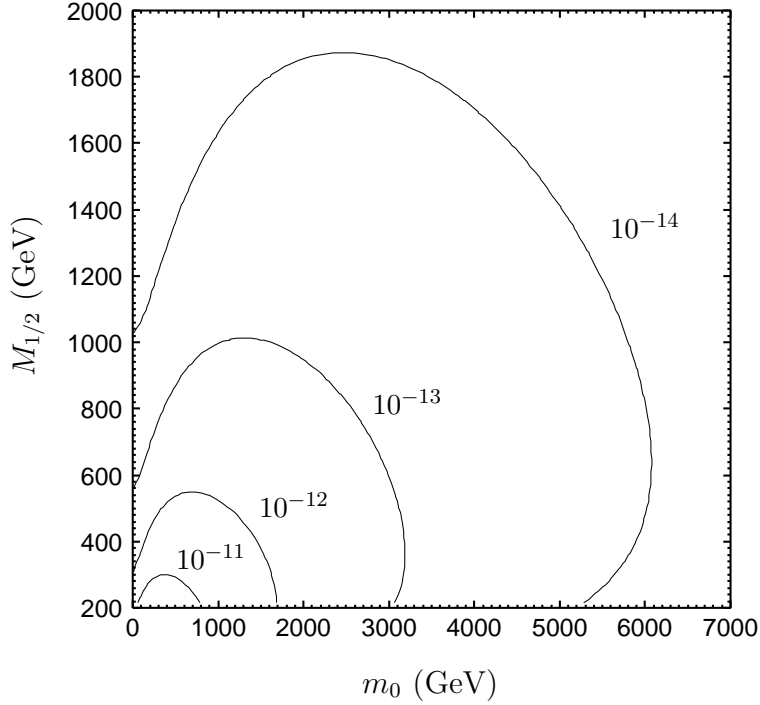}  
\end{center}
\caption{
The dependence of $B(\mu \to e \gamma)$ on the SUSY breaking parameters
are plotted.
We take $\tilde{f}_\nu^{33} = 1/\sqrt{2}$ and
the neutrino oscillation parameters of
the large mixing MSW solution and $U^{e3}_{\rm MNS} = 0$.
The horizontal and vertical axes represent
the universal scalar mass and the gaugino mass 
given at the Planck scale.
The lines represent
the branching ratio of 
$10^{-11}$, $10^{-12}$, $10^{-13}$, and $10^{-14}$
from inside.
The value of the $\tan \beta$, which is the ratio of the 
two VEVs of the Higgs fields,
is taken to be $\tan \beta = 10$.
\label{fig2}
}
\end{figure}

As a prediction of our model,
we consider LFV processes such as
the $\mu \to e \gamma$ and $\tau \to \mu \gamma$ decays.
An interesting point of the Dirac neutrino
is that the Yukawa coupling constants are
directly related to the neutrino oscillation parameters,
namely
\begin{eqnarray}
 f_\nu^{ij} \propto U_{\rm MNS}^{ij} m_{\nu j}\ ,
\label{6}
\end{eqnarray}
where $U_{\rm MNS}$ and $m_\nu$ are the Maki-Nakagawa-Sakata (MNS)
matrix \cite{Maki:1962mu} and the neutrino masses.
In this case,
the branching ratios of the LFV processes are
strongly correlated with the neutrino oscillation parameters.
In the minimal supergravity scenario,
in which the slepton mass matrix is 
proportional to the unit matrix at the tree level,
off-diagonal components of the slepton matrix
are induced through the loop diagrams
with the LFV interactions
\cite{Hall:1986dx}.
In the seesaw model,
the relation between the neutrino oscillation parameters
and the LFV processes have been investigated in detail 
\cite{Borzumati:1986qx, Casas:2001sr}.
In that case,
we need to assume the mass matrix of the right-handed neutrinos 
and the results depend on the pattern of 
the neutrino mass matrix, i.e.,
hierarchical or degenerate \cite{Casas:2001sr}.
However,
in our case,
the prediction is directly related to the
observables in the neutrino oscillation experiments,
i.e., mass squared differences and mixing angles
as we see in the following.

In the minimal supergravity scenario,
the off-diagonal components 
of the left-handed slepton mass matrix
are induced 
by the renormalization group running between
the Planck scale to the GUT scale
through the $\bar{F} H^\prime N$ interactions,
and are approximately given by
\begin{eqnarray}
 (\tilde{m}^2_{\tilde{l}})_{ij} \simeq
- \frac{1}{8 \pi^2} \sum_{k}
\tilde{f}_\nu^{ik*} \tilde{f}_\nu^{jk}
(3 + |a_0|^2) m_0^2 \log \frac{M_{\rm GUT}}{M_{\rm Pl}}\ ,
\label{7}
\end{eqnarray}
where $a_0$ and $m_0$ are 
the coupling constant of the universal three point scalar interaction
and the universal scalar masses, respectively.
From eqs.(\ref{6}) and (\ref{7}),
we can describe the off-diagonal components as follows:
\begin{eqnarray}
 (\tilde{m}^2_{\tilde{l}})_{e \mu} \propto
U_{\rm MNS}^{e1*} U_{\rm MNS}^{\mu 1} \Delta m^2_{12} +
U_{\rm MNS}^{e3*} U_{\rm MNS}^{\mu 3} \Delta m^2_{32} 
\ ,
\label{8}
\end{eqnarray}
\begin{eqnarray}
 (\tilde{m}^2_{\tilde{l}})_{\mu \tau} \propto
U_{\rm MNS}^{\mu 1*} U_{\rm MNS}^{\tau 1} \Delta m^2_{12} +
U_{\rm MNS}^{\mu 3*} U_{\rm MNS}^{\tau 3} \Delta m^2_{32} 
\ ,
\label{9}
\end{eqnarray}
\begin{eqnarray}
 (\tilde{m}^2_{\tilde{l}})_{e \tau} \propto
U_{\rm MNS}^{e 1*} U_{\rm MNS}^{\tau 1} \Delta m^2_{12} +
U_{\rm MNS}^{e 3*} U_{\rm MNS}^{\tau 3} \Delta m^2_{32} 
\ ,
\label{10}
\end{eqnarray}
where $\Delta m^2_{ij} = m_{\nu i}^2 - m_{\nu j}^2$.
From these equations,
we can predict the ratio of the branching ratios
such as $B(\tau \to \mu \gamma)/ B(\mu \to e \gamma)$
once the neutrino oscillation parameters are determined.
For example,
in case of the large mixing MSW solution
for the solar neutrino problem \cite{Wolfenstein:1978ue}
and 
$U_{\rm MNS}^{e3} = 0$,
the ratio of the branching ratios of 
$\mu \to e \gamma$ and $\tau \to \mu \gamma$ is 
given by
\begin{eqnarray}
 \frac{B(\tau \to \mu \gamma)}{B(\mu \to e \gamma)}
\simeq 0.35 \left( \frac{(\tilde{m}^2_{\tilde{l}})_{\mu \tau}}
{(\tilde{m}^2_{\tilde{l}})_{e \mu}} 
\right)^2
\simeq 0.35 \left( \frac{\Delta m^2_{32}}{\Delta m^2_{12}} \right)^2
\simeq 10^{3}
\ .
\label{11}
\end{eqnarray}
In GUT models,
there is another contribution
to the off-diagonal components of the 
right-handed slepton mass matrix
through the LFV interaction
between the right-handed quarks and leptons
with the CKM mixing \cite{Barbieri:1994pv}.
However,
those contributions are negligibly small compared
to those from $\bar{F} H^\prime N$ interactions
for $\tilde{f}_\nu^{33} \sim 1$.
For the case of large and small angle MSW solutions,
we plot the ratio in Fig.\ref{fig1}
as a function of $U^{e3}_{\rm MNS}$.
We can see significant dependence on $U^{e3}_{\rm MNS}$.
The ratio varies
from 10 to 2000 for large angle MSW solution and 
from 10 to 30000 for small angle MSW solution.
Dependence of $B(\mu \to e \gamma)$ on the SUSY breaking
parameters are given in Fig.\ref{fig2}.
We take the coupling constant $\tilde{f}_\nu^{33} = 1/\sqrt{2}$
and neutrino oscillation parameter of the large angle MSW solution
and $U^{e3}_{\rm MNS} = 0$.
The parameter $m_0$ and $M_{1/2}$ are 
the universal scalar mass and the gaugino mass
given at the Planck scale.
In the wide region of the SUSY breaking parameters,
$B(\mu \to e \gamma)$ is greater than $10^{-14}$
which is within the reach of the planed experiments 
at PSI \cite{PSI} and JHF \cite{PRISM}.

In conclusion,
we proposed a simple mechanism to generate
Dirac neutrino masses in SU(5) SUSY GUT
in which the tiny coupling constants are induced by
the mixing between the Higgs field and 
another superheavy Higgs field.
The tiny mixing is 
realized by the hierarchy of the mass parameter
$M_{H^\prime} \sim M_{\rm GUT}$ and 
$\mu^\prime \sim M_{\rm SUSY} \sim$ TeV,
and this situation is exactly 
the same as the $\mu$-problem \cite{Kim:1984dt}.
Therefore using the same mechanism for solving the $\mu$-problem,
we can obtain such hierarchy and the tiny Dirac neutrino masses
are naturally explained.
We emphasize that the small Dirac neutrino masses
are obtained in SUSY GUT 
without introducing any new scale parameters.
As an interesting feature of this model,
the branching ratios of the LFV processes
are directly related to the neutrino oscillation parameters.
We calculated the branching ratios and found that
$B(\mu \to e \gamma)$ is large enough to be observable
and the ratio $B(\tau \to \mu \gamma)/B(\mu \to e \gamma)$
is predicted once the neutrino oscillation parameters
are determined.

Finally, we would like to comment on an alternative model.
If we add three pairs of 
$F^\prime:({\bf 5}, -1)$ and 
$\bar{F}^\prime:({\bf \bar{5}}, 3)$
instead of the extension of the Higgs sector,
the superpotential is given by
\begin{eqnarray}
 W = {{y}}_\nu^{ij} \bar{F}^\prime_i H N_j
+ M_{F^\prime}^{ij} F^\prime_i \bar{F}^\prime_j
+ \mu H \bar{H}
+ \tilde{\mu}_{ij} F^\prime_i \bar{F}_j\ .
\end{eqnarray}
By diagonalizing the mass matrix,
we obtain small Dirac neutrino masses.
However, in this case,
the direct relation eq.(\ref{8}--\ref{10}) are lost
because the low energy 
neutrino Yukawa coupling constants $f_\nu$ are not proportional
to the original one ${{y}}_\nu$ in general.

\section*{Acknowledgments}
We would like to thank K.~Hamaguchi, T.~Moroi, and K.~Yoshioka
for useful discussions.
This work was supported by the JSPS Research Fellowships
for Young Scientists.
%


\end{document}